\documentclass[aps,prd,twocolumn,nofootinbib,showpacs]{revtex4-1}

\usepackage{graphicx,epsfig,color}

\usepackage[colorlinks=true,linkcolor=blue,citecolor=blue, urlcolor=blue]{hyperref}

\begin{document}

\title{Reanalysis of the Higgs-boson decay  $H \to gg$ up to $\alpha_s^6$-order level using the principle of maximum conformality}

\author{Jun Zeng}
\author{Xing-Gang Wu} \email{wuxg@cqu.edu.cn}
\author{Shi Bu}
\author{Jian-Ming Shen}
\affiliation{Department of Physics, Chongqing University, Chongqing 401331, P.R. China}

\author{Sheng-Quan Wang}
\affiliation{School of Mechatronics Engineering, Guizhou Minzu University, Guiyang 550025, P.R. China}

\begin{abstract}

Using the newly available $\alpha_s^6$-order QCD correction to the Higgs decay channel $H\to gg$, we make a detailed discussion on the perturbative properties of the decay width $\Gamma(H\to gg)$ by using the principle of maximum conformality (PMC). The PMC provides a way to eliminate the conventional renormalization scheme-and-scale ambiguities, which uses the renormalization group equation to determine the optimal running behavior of the strong coupling constant at each order via a recursive way. Even though there is no ambiguity for setting the renormalization scale, there is residual scale dependence for the PMC predictions due to unknown high-order terms. Using the $\alpha_s^6$-order terms, the somewhat larger residual renormalization scale dependence at the $\alpha_s^5$-order level observed in our previous work can be greatly suppressed, which shows $\Gamma (H\to gg)\rm{|_{ PMC }} =337.9 \pm1.7_{-0.1}^{+0.9}\pm1.9$ KeV, where the first error is caused by the Higgs mass uncertainty $\Delta M_{H}=0.24$ GeV, the second one is the residual scale dependence by varying the initial choice of scale within the region of $\left[{M_H}/{2},4 M_H\right]$, and the third one is the conservative prediction of unknown high-order contributions.

\end{abstract}


\maketitle

\section{Introduction}

The Higgs boson is an important component of the Standard Model (SM), which arouses people's great interest either for precision test of the SM or for searching of new physics beyond the SM. Among its decay channels, the $H\to gg$ decay plays an important role in Higgs phenomenology. At present, the next-to-leading order $(\rm{NLO})$~\cite{Inami:1982xt, Djouadi:1991tka, Graudenz:1992pv, Dawson:1993qf, Spira:1995rr, Dawson:1991au}, the next-to-next-to-leading order $(\rm{N^{2}LO})$~\cite{Chetyrkin:1997iv, Chetyrkin:1997un}, the next-to-next-to-next-to-leading order $(\rm{N^{3}LO})$~\cite{Baikov:2006ch}, and the next-to-next-to-next-to-next-to-leading order $(\rm{N^{4}LO})$~\cite{Herzog:2017dtz} perturbative Quantum Chromodymaics (pQCD) corrections to the Higgs decay width $\Gamma  (H\to gg)$ have been given in the literature. Those achievements, especially the newly achieved state-of-the-art $\rm{N^{4}LO}$-term, provide us with a great opportunity for achieving precise pQCD predictions.

Because of renormalization group invariance, the physical observable should be independent of the choices of renormalization scheme and renormalization scale. However for a fixed-order prediction of the observable, the mismatch of the strong coupling constant and the pQCD coefficients at each perturbative order leads to the well-known renormalization scheme-and-scale ambiguities. Conventionally, one chooses the ``guessed" typical momentum flow of the process or the one to eliminate the large logs as the renormalization scale with the purpose of minimizing such scale dependence around its central point. One hopes to achieve a small scheme-and-scale dependent prediction by finishing higher-and-higher order QCD corrections.

Guessing the renormalization scale and setting an arbitrary range for its value introduces an arbitrary systematic error into pQCD predictions, which may lead to predictions inconsistent with the experimental measurements. It is helpful to have a guiding principle for setting the renormalization scale to fix such kind of problems. The principle of maximum conformality (PMC)~\cite{Brodsky:2011ta, Brodsky:2012rj, Mojaza:2012mf, Brodsky:2013vpa} has been suggested for the purpose of eliminating the conventional renormalization scheme-and-scale ambiguities. It has been found that the PMC satisfies all the self-consistency conditions of the renormalization group equation (RGE) or the $\beta$-function~\cite{Brodsky:2012ms} and the renormalization group invariance, which has been applied for various processes, c.f. the reviews~\cite{Wu:2013ei, Wu:2014iba, Wu:2015rga}.

The essential PMC procedure is to identify all contributions which originate from the $\beta$-terms in a pQCD series; one then shifts the renormalization scale of the QCD running coupling at each order to absorb those $\beta$-terms. The coefficients of the resulting scale-fixed perturbative series is thus identical to the coefficients of the scheme-independent ``conformal" series. New $\beta$-terms will occur at each order, so the PMC scale for each order is generally distinct, which reflects the varying virtuality of the amplitude that occurs at each order. One may choose any value (in perturbative region) as the initial renormalization scale, and the determined PMC scale, corresponding to the correct running behavior of the strong coupling at this particular order, shall be highly independent of such choice, thus solving the conventional renormalization scale ambiguity.

The accuracy of the PMC prediction depends heavily on how well we know the scale-running behavior of the strong coupling constant $\alpha_s$, which could be determined by using the RGE via a superposition way~\cite{Mojaza:2012mf, Brodsky:2013vpa}. The conventional RGE can be solved recursively, whose solution is now known up to five-loop level~\cite{Brodsky:2011ta, Kniehl:2006bg, Baikov:2016tgj, Tarasov:1980au, Larin:1993tp, vanRitbergen:1997va, Chetyrkin:2004mf, Czakon:2004bu, Herzog:2017ohr}. In those solutions, the QCD asymptotic scale $\Lambda$ can be determined by using the world average of $\alpha_s$ at the scale $M_Z$, e.g. $\alpha^{\rm exp}_s(M_Z)=0.1181\pm 0.0011$, which leads to $\Lambda^{n_f=5}_{\overline{\rm MS}}=0.210\pm0.014$ GeV~\cite{Patrignani:2016xqp}. The asymptotic scale under different renormalization scheme can be transformed by Celmaster-Gonsalves relation~\cite{Celmaster:1979km, Celmaster:1979dm, Celmaster:1979xr, Celmaster:1980ji}. The $\{\beta_i\}$-functions are generally expressed by the $n_f$-power series, $n_f$ being the active flavor number. In dealing with the pQCD approximate, we can first transform the usual $n_f$-power series at each order into the $\{\beta_i\}$-series and then apply the standard PMC procedures. It should be noted that only those $\{\beta_i\}$-terms that are pertained to the renormalization of running coupling should be absorbed into the running coupling so as to achieve the optimal scales of the process. For example, the light-by-light quark loop term is proportional to $n_f$; since it is free of ultraviolet divergences, it should be treated as a ``conformal" contribution when applying the PMC. Then, special treatment should be paid for distributing the $\{\beta_i\}$-terms of the process. The physical momentum space subtraction scheme (mMOM-scheme)~\cite{Celmaster:1979km, Celmaster:1979dm, Celmaster:1979xr, Celmaster:1980ji, Gracey:2013sca, vonSmekal:2009ae} carries information of the vertex at specific momentum configuration. This external momentum configuration is non-exceptional and there are no infrared issues, thus avoiding the confusion of distinguishing $\{\beta_i\}$-terms. Thus to avoid the complexity of applying the PMC within the $\overline{\rm MS}$-scheme, similar to the case of QCD BFKL Pomeron~\cite{Brodsky:1998kn, Zheng:2013uja, Hentschinski:2012kr, Caporale:2015uva}, one can first transform the results from the $\overline{\rm MS}$-scheme to the and then apply the PMC scale setting. The MOM-scheme is gauge dependent, and for definiteness, we shall adopt the Landau gauge to do our calculation.

Refs.\cite{Wang:2013bla, Zeng:2015gha} have presented a PMC analysis on the Higgs decay width $\Gamma  (H\to gg)$ up to $\alpha_s^5$-order level. It shows a good application of PMC, the residual scale dependence for the total decay width and the decay widths of most of the separate orders are negligibly small. An exception is that the PMC scale for the $\rm{NLO}$-term of the decay width $\Gamma_{\rm{NLO}}$ has poor convergence, which leads to a somewhat larger residual scale dependence~\cite{Zeng:2015gha}. It is interesting to know whether such large residual scale dependence can be suppressed by using the newly achieved $\alpha_s^6$-terms, as is the purpose of the present paper. Moreover, since all the PMC scales shall be improved by the new high-order terms, a more accurate prediction on the decay width $\Gamma  (H\to gg)$ can be achieved.

The PMC scale is in pQCD series, whose perturbative coefficients are determined by the non-conformal $\{\beta_i\}$-terms of pQCD series. There is residual scale dependence in the finite-order PMC predictions, which is caused by the unknown high-order terms of the PMC scale, especially we have no $\{\beta_i\}$-term information to set the PMC scale of the highest perturbative order. This residual scale dependence is quite different from the above mentioned conventional scale-setting uncertainty, which is purely a guess work. As has been observed in many of PMC applications, such residual uncertainties are highly suppressed, even for low-order predictions, due to the rapid convergence of the conformal pQCD series.

The remaining parts of this paper are organized as follows. We will give the PMC analysis of the decay width $\Gamma(H\to gg)$ up to $\alpha_s^6$-order level in Sec.II. Numerical results are given in Sec.III. Sec.IV is reserved for a summary. For convenience, we present the new coefficients emerged at the $\alpha_s^6$-order level in the Appendix.

\section{The PMC analysis of the decay width $\Gamma(H\to gg)$ up to $\alpha_s^6$-order level}

Up to $\alpha_s^6$-order level, the decay width of $H\to gg$ takes the form
\begin{equation}
\Gamma  (H\to gg)=\frac{M_H^3 G_F}{36 \sqrt{2} \pi }\sum _{k=0}^{4 }  C_{k}(\mu_r) a^{k+2}_{s}(\mu_r),   \label{rij}
\end{equation}
where $a_s=\alpha_s/4\pi$ and $\mu_r$ stands for the arbitrary initial choice of renormalization scale. The perturbative coefficients, $ C_{k\in[0,4]}(M_H)=\sum _{j=0}^k c_{k+1,j}(M_H) n_f^j$, whose expressions under the $\rm{\overline {MS}}$-scheme can be read from Ref.\cite{Herzog:2017dtz}. We can conveniently get their values at any other scale by using the RGE via a recursive way.

As discussed in the Introduction, we need to transform the $\overline{\rm MS}$-scheme pQCD series into mMOM one, which can be achieved by using the newly available relationship between the $\rm{\overline {MS}}$-scheme strong coupling and the mMOM-scheme strong coupling up to $\rm{N^4LO}$-level~\cite{Ruijl:2017eht}. More explicitly, under the Landau gauge, their relations up to $\rm{N^4LO}$-level are
\begin{equation}
a_{s}(\mu_r)=\sum_{i=1}^{5} d^{(i)} a^i_{s,\rm{\overline {MS}}}(\mu_r),
\end{equation}
where $a_{s}$ stands for the MOM-coupling and the expansion coefficients $d^{(i)}$ are given by
\begin{eqnarray}
d^{(1)}=&&1, \\
d^{(2)}=&&\frac{169}{12}-\frac{10}{9} n_f, \\
d^{(3)}=&&-\frac{351 \zeta_{3}}{8}+\frac{76063}{144}- \bigg(\frac{4 \zeta_{3}}{3}+\frac{1913}{27}\bigg) n_f +\frac{100}{81} n_f^2,  \\
d^{(4)}=&&-\frac{60675 \zeta_{3}}{16}-\frac{70245 \zeta_{5}}{64}+\frac{42074947}{1728}\nonumber \\
&&+\bigg(\frac{8362 \zeta_{3}}{27}+\frac{2320 \zeta_{5}}{9}-\frac{769387}{162}\bigg)n_f\nonumber \\
&&+\left(\frac{28 \zeta_{3}}{9}+\frac{199903}{972}\right) n_f^2-\frac{1000}{729} n_f^3,
\end{eqnarray}
\begin{widetext}
\begin{eqnarray}
d^{(5)}=&&-\frac{13941 \zeta_{3}^2}{32}-\frac{139007835 \zeta_{3}}{512}+\frac{655135047 \zeta_{7}}{4096}-\frac{174626085 \zeta_{5}}{512}+\frac{297 \pi^{4}}{20}+\frac{14543057783}{10368}\nonumber \\
&&
+\bigg(\frac{421 \zeta_{3}^2}{8}+\frac{18577535 \zeta_{3}}{432}+\frac{17195455 \zeta_{5}}{216}-\frac{26952037 \zeta_{7}}{864}-\frac{1627 \pi^{4}}{1620}-\frac{2772344147}{7776}\bigg)n_f\nonumber \\
&&
+\bigg(\frac{400 \zeta_{3}^2}{9}-\frac{201631 \zeta_{3}}{324}-\frac{1088305 \zeta_{5}}{324}+\frac{809 \pi^{4}}{2430}+\frac{16236365}{648}\bigg)n_f^2\nonumber \\
&&
+\bigg(-\frac{284 \zeta_{3}}{27}+\frac{880 \zeta_{5}}{27}-\frac{8919413}{17496}\bigg)n_f^3+\frac{10000}{6561}n_f^4,
\end{eqnarray}
\end{widetext}
where $\zeta_{n}$ are Riemannian Zeta functions. We can then obtain the coefficients $c_{i,j}$ under the MOM-scheme. The coefficients up to $\alpha_s^5$-order level has been given in Ref.\cite{Zeng:2015gha}, and for convenience, we present the new ones $c_{5,j}$ at the $\alpha_s^6$-order in the Appendix.

The $\beta_{i}$-functions under the mMOM scheme up to five-loop level can be found in Ref.\cite{Ruijl:2017eht},
\begin{eqnarray}
\beta_0=&&11-\frac{2}{3}n_f,\\
\beta_1=&&102-\frac{38}{3}n_f,\\
\beta_2=&&-\frac{3861 \zeta_{3}}{8}+\frac{28965}{8}+\left(\frac{175 \zeta_{3}}{12}-\frac{7715}{12}\right)n_f\nonumber \\
&&+\left(\frac{8 \zeta_{3}}{9}+\frac{989}{54}\right)n_f^2,\\
\beta_3=&&-\frac{625317 \zeta_{3}}{16}-\frac{772695 \zeta_{5}}{32}+\frac{1380469}{8}\nonumber \\
&&+\left(\frac{516881 \zeta_{3}}{72}+\frac{1027375 \zeta_{5}}{144}-\frac{970819}{24}\right)n_f\nonumber \\
&&+\bigg(\frac{736541}{324}-\frac{6547 \zeta_{3}}{27}-\frac{9280 \zeta_{5}}{27}\bigg) n_f^2\nonumber \\
&&+\bigg(\frac{16 \zeta_{3}}{9}-\frac{800}{27}\bigg)n_f^3,
\end{eqnarray}
\begin{widetext}
\begin{eqnarray}
\beta_4=&&-\frac{7696161 \zeta_{3}^2}{64}-\frac{1064190195 \zeta_{3}}{512}+\frac{21619456551 \zeta_{7}}{4096}-\frac{4922799165 \zeta_{5}}{512}+\frac{3248220045}{256}\nonumber \\
&&+\bigg(\frac{82869 \zeta_{3}^2}{32}+\frac{10327103555 \zeta_{3}}{20736}+\frac{18219328375 \zeta_{5}}{6912}-\frac{24870449471 \zeta_{7}}{18432}-\frac{115659378547}{31104}\bigg)n_f\nonumber \\
&&+\bigg(\frac{59531 \zeta_{3}^2}{36}-\frac{13019053 \zeta_{3}}{1296}+\frac{26952037 \zeta_{7}}{432}-\frac{65264845 \zeta_{5}}{324}+\frac{833934985}{2592}\bigg)n_f^2\nonumber \\
&&+\bigg(-\frac{2240 \zeta_{3}^2}{27}-\frac{129869 \zeta_{3}}{162}+\frac{299875 \zeta_{5}}{54}-\frac{3249767}{324}\bigg) n_f^3\nonumber \\
&&+\bigg(\frac{304 \zeta_{3}}{27}-\frac{1760 \zeta_{5}}{27}+\frac{2617}{27}\bigg) n_f^4.
\end{eqnarray}
\end{widetext}
We can then transform the $n_f$-series into the $\{\beta_i\}$-series, i.e.
\begin{widetext}
\begin{eqnarray}
\Gamma  (H\to gg)=&&\frac{M_H^3 G_F}{36 \sqrt{2} \pi }\bigg[ r_{1,0}a_{s}^2(\mu_r)+\left(r_{2,0}+2 r_{2,1} \beta _0 \right) a_{s}^3(\mu_r)+ \left(r_{3,0}+3 r_{3,1} \beta _0 +3 r_{3,2} \beta_0^2+2 r_{2,1} \beta _1\right) a_{s}^4(\mu_r) \nonumber \\
&&+\left(r_{4,0}+7 r_{3,2} \beta _1 \beta _0 +4 r_{4,1} \beta _0+6 r_{4,2} \beta_0^2+4 r_{4,3} \beta_0^3 +2 r_{2,1} \beta _2 +3 r_{3,1} \beta _1 \right) a_{s}^5(\mu_r)  \nonumber \\
&&+ \left(r_{5,0}+8 r_{3,2} \beta _2  \beta _0 +\frac{27}{2} r_{4,2} \beta _1  \beta _0 +5 r_{5,1} \beta _0  +\frac{47}{3} r_{4,3} \beta_1  \beta _0^2+10 r_{5,2} \beta_0^2+10 r_{5,3} \beta_0^3  \right. \nonumber \\
&& \left. +5 r_{5,4} \beta_0^4 +2 r_{2,1} \beta _3 +3 r_{3,1} \beta_2 +4 r_{3,2} \beta_1^2 +4 r_{4,1} \beta_1\right) a_{s}^6(\mu_r) \bigg]+\mathcal{O}\left(a_{s}^7(\mu_r)\right),  \label{betarij}
\end{eqnarray}
\end{widetext}
where the coefficients $r_{i,j}$ up to $\alpha_s^5$-order level has been given in Ref.\cite{Zeng:2015gha}, and for convenience, we present the new ones $c_{5,j}$ at the $\alpha_s^6$-order in the Appendix.

Following the standard PMC procedures, we can get the optimal behavior of the running coupling at each order up to $\alpha_s^6$-order level. After applying the PMC, the pQCD series (\ref{betarij}) can be rewritten as the following scheme-independent conformal series
\begin{equation}
\Gamma  (H\to \ gg) = \frac{M_H^3 G_F}{36 \sqrt{2} \pi } \sum _{j=1}^{5 }  r_{j,0} a_{s}^{j+1}(Q_j),
\end{equation}
where $r_{j,0}$ are conformal coefficients, which are free of the non-conformal $\{\beta_i\}$-terms. The PMC scales $Q_j$ can be written as
\begin{eqnarray}
\ln{\frac{Q_1^2}{\mu_r^2}}&=&-\frac{r_{2,1}}{r_{1,0}}+\frac{3}{2 r_{1,0}^2}  \left(r_{2,1}^2-r_{1,0} r_{3,2}\right) \beta_{0} a_s(\mu_r) \nonumber \\
&&+ \frac{1}{2 r_{1,0}^3} \bigg[\left(-5 r_{2,1}^3+9 r_{1,0} r_{3,2} r_{2,1}-4 r_{1,0}^2 r_{4,3}\right)  \beta_0^2 \nonumber \\
&& +4 r_{1,0} \left(r_{2,1}^2-r_{1,0} r_{3,2}\right) \beta_1\bigg] a^2_{s}(\mu_r)  \nonumber \\
&&+ \frac{1}{24 r_{1,0}^4} \bigg \{\beta_0 \bigg[3 \left(35 r_{2,1}^4-90 r_{1,0} r_{3,2} r_{2,1}^2 \right.  \nonumber \\
&&\left.+48 r_{1,0}^2 r_{4,3} r_{2,1}+27 r_{1,0}^2 r_{3,2}^2-20 r_{1,0}^3 r_{5,4}\right)  \beta_0^2  \nonumber \\
&& -4 r_{1,0} \left(37 r_{2,1}^3-72 r_{1,0} r_{3,2} r_{2,1}+35 r_{1,0}^2 r_{4,3}\right) \beta_1\bigg ]  \nonumber \\
&&-60 r_{1,0}^2 \left(r_{1,0} r_{3,2}-r_{2,1}^2\right) \beta_2 \bigg\} a^3_{s}(\mu_r) ,  \label{PMCQ11} \\
\ln{\frac{Q_2^2}{\mu_r^2}}&=&-\frac{r_{3,1}}{r_{2,0}}+\frac{2}{r_{2,0}^2}  \left(r_{3,1}^2-r_{2,0} r_{4,2}\right) \beta_0 a_{s}(\mu_r)\nonumber \\
&& -\frac{1}{6 r_{2,0}^3} \bigg
[4 \left(7 r_{3,1}^3-12 r_{2,0} r_{4,2} r_{3,1}+5 r_{2,0}^2 r_{5,3}\right) \beta_0^2  \nonumber \\
&& +15 r_{2,0} \left(r_{2,0} r_{4,2}-r_{3,1}^2\right) \beta_1  \bigg ] a^2_{s}(\mu_r) , \label{PMCQ12} \\
\ln{\frac{Q_3^2}{\mu_r^2}}&=&-\frac{r_{4,1}}{r_{3,0}}+\frac{5}{2 r_{3,0}^2}  \left(r_{4,1}^2-r_{3,0} r_{5,2}\right) \beta_0 a_{s}(\mu_r) ,  \label{PMCQ13} \\
\ln{\frac{Q_4^2}{\mu_r^2}}&=& -\frac{r_{5,1}}{r_{4,0}}. \label{PMCQ14}
\end{eqnarray}
The PMC scales at different orders are of different accuracy, which are determined by iteratively using the RGE and resum different $\{\beta_i\}$-terms into their perturbative series~\cite{Mojaza:2012mf, Brodsky:2013vpa}. By using the perturbative series of $\Gamma(H\to gg)$ up to $\alpha_s^6$-level, we can determine the LO PMC scale $Q_1$ up to next-to-next-to-next-leading logarithmic order $({\rm N^{3}LLO})$ accuracy,  the NLO PMC scale $Q_2$ up to next-to-next-leading logarithmic order $({\rm N^{2}LLO})$ accuracy, the ${\rm N^2 LO}$ PMC scale $Q_3$ up to next-leading logarithmic order $({\rm NLLO})$ accuracy, and the ${\rm N^3 LO}$ PMC scale $Q_4$ at the leading logarithmic order $({\rm LLO})$ accuracy, respectively. To compare with the PMC scales $Q_i$ for a $\alpha_s^5$-order prediction~\cite{Zeng:2015gha}, the accuracy of the PMC scales have been improved by the new $\{\beta_i\}$-terms emerged at the $\alpha_s^6$-order level, i.e. their highest-order terms are determined by the $\alpha_s^6$-order terms. As for $Q_5$ which needs the $\{\beta_i\}$-term information at the uncalculated $\alpha_s^7$-order level and is undetermined, our optimal choice for $Q_5$ is the last known PMC scale $Q_4$, which ensures the scheme-independence of the resultant PMC pQCD series~\cite{Mojaza:2012mf, Brodsky:2013vpa}.

\section{Numerical results}

To do the numerical calculation, we take $G_F=1.16638 \times 10^{-5}~{\rm{GeV}}^{-2}$, the top-quark pole mass $m_t = 173.3~\rm{GeV}$~\cite{ACCPT:2012TOP}, and $M_H = 125.09 \pm 0.21 \pm 0.11$~$\rm{GeV}$~\cite{Aad:2015zhl}.

\subsection{Perturbative nature of the decay width $\Gamma (H \to gg)$ up to $\rm{N^4LO}$ level}

\begin{figure}[htb]
\centering
\includegraphics[width=0.470\textwidth]{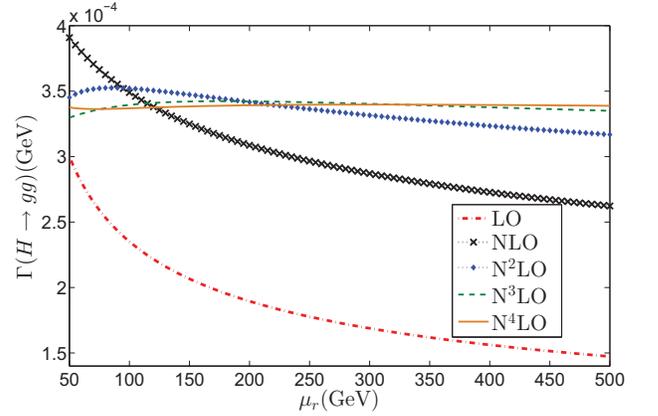}
\caption{Total decay width $\Gamma (H \to gg)$ under conventional scale setting. The dash-dot line, the dotted line with cross symbols, the dotted lines with rhombus symbols, the dashed line and the solid line are for the predictions up to $\rm{LO}$, $\rm{NLO}$, $\rm{N^2LO}$, $\rm{N^3LO}$, and $\rm{N^4LO}$ levels, respectively.}
\label{HtoggConv}
\end{figure}

\begin{figure}[htb]
\centering
\includegraphics[width=0.470\textwidth]{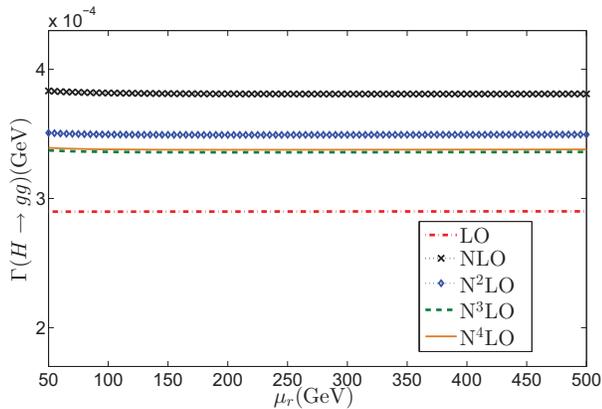}
\caption{Total decay width $\Gamma (H \to gg)$ after applying the PMC.  The dash-dot line, the dotted with cross symbols, the dotted line with rhombus symbols, the dashed line and the solid line are for the predictions up to $\rm{LO}$, $\rm{NLO}$, $\rm{N^2LO}$, $\rm{N^3LO}$, and $\rm{N^4LO}$ levels, respectively.}
\label{mMOMPMC}
\end{figure}

We present the total decay width $\Gamma (H \to gg)$ up to $\rm{N^4LO}$ level before and after applying the PMC in Figs.(\ref{HtoggConv}, \ref{mMOMPMC}). When more loop terms have been taking into consideration, the conventional scale dependence becomes smaller. Up to N$^4$LO-level, the total decay width under conventional scale-setting is almost flat versus the initial choice of scale. As a comparison, the PMC prediction for the total decay width is scale-independent even for low-order predictions. This shows that if one can determine the correct behavior of the running coupling, one can get the scale-independent prediction at any fixed order.

\begin{table}[htb]
\begin{center}
\begin{tabular}{  c c c  c  c c c c c }
\hline
& ~$ \Gamma_i$~          & ~$\rm{LO}$~        & ~$\rm{NLO}$~    & ~$\rm{N^2LO}$~        & ~$\rm{N^3LO}$~      & ~$\rm{N^4LO}$~    & ~$\rm{Total}$~       \\
\hline
& $\Gamma_i |_{M_H/2}$  & 276.22   &   101.32  & $-$27.33  & $-$16.85 &3.11 &336.48    \\
& $\Gamma_i |_{M_H}$  & 218.69& 116.86  &14.48 & $-$8.76 & $-$3.60& 337.67  \\
& $\Gamma_i |_{2M_H}$  &   177.65    &   118.87      &     39.78   &    5.16      &    $-$1.84   &     339.63   \\
& $\Gamma_i |_{4M_H}$  &   147.31    &   114.91      &   54.58     &   18.18    &     3.85 &     338.83   \\
\hline
\end{tabular}
\caption{Total and individual decay widths (in unit: KeV) of the decay $H \to gg$ under conventional scale-setting. $\Gamma_i$ stands for the individual decay
width at each order with $i =\rm{LO}$, $\rm{NLO}$, $\rm{N^2LO}$, $\rm{N^3LO}$, and $\rm{N^4LO}$, respectively. $\rm{\Gamma_{Total}}=\sum \Gamma_i$ stands for total decay width. Several typical scales, $\mu_r = M_H/2$, $M_H$, $2M_H$, and $4M_H $, are adopted.}
\label{conv}
\end{center}
\end{table}

\begin{table}[htb]
\begin{center}
\begin{tabular}{  c c c  c  c c c c c }
\hline
& ~$ \Gamma_i$~          & ~$\rm{LO}$~        & ~$\rm{NLO}$~    & ~$\rm{N^2LO}$~        & ~$\rm{N^3LO}$~      & ~$\rm{N^4LO}$~    & ~$\rm{Total}$~     \\
\hline
& $\Gamma_i |_{M_H/2}$  & 289.84   &   92.77  & $-$32.36  & $-$13.47 &1.95 &338.72    \\
& $\Gamma_i |_{M_H}$  & 289.85& 91.51 &$-$31.97 &$-$13.47 & 1.95  & 337.86   \\
& $\Gamma_i |_{2M_H}$  &   289.93    &   90.99     &     $-$31.67   &    $-$13.47     &     1.95   &     337.73   \\
& $\Gamma_i |_{4M_H}$  &   290.07    &   90.87       &    $-$31.42    &   $-$13.47      &     1.95     &    338.00   \\
\hline
\end{tabular}
\caption{Total and individual decay widths (in unit: KeV) of the decay $H \to gg$ under PMC scale-setting. $\Gamma_i$ stands for the individual decay
width at each order with $i =\rm{LO}$, $\rm{NLO}$, $\rm{N^2LO}$, $\rm{N^3LO}$, and $\rm{N^4LO}$, respectively. $\rm{\Gamma_{Total}}=\sum \Gamma_i$ stands for total decay width. Several typical scales, $\mu_r = M_H/2$, $M_H$, $2M_H$, and $4M_H $, are adopted.}
\label{pmcdw}
\end{center}
\end{table}

We present the total and the individual decay widths, $\Gamma_{\rm Total}$ and $\Gamma_i$, of the decay $H \to gg$ under conventional and PMC scale-setting approaches in Tables \ref{conv} and \ref{pmcdw}, where $\rm{\Gamma_{Total}}=\sum \Gamma_i$ with $i =\rm{LO}$, $\rm{NLO}$, $\rm{N^2LO}$, $\rm{N^3LO}$, and $\rm{N^4LO}$, respectively. For conventional scale-setting,
\begin{equation}
\Gamma_{\rm{N^kLO}}=\frac{M_H^3 G_F}{36 \sqrt{2} \pi} C_{k}(\mu_r) a_{s}^{k+2}(\mu_r),
\end{equation}
and for PMC scale-setting
\begin{equation}
\Gamma_{\rm{N^kLO}}=\frac{M_H^3 G_F}{36 \sqrt{2} \pi } r_{k+1,0}a_{s}^{k+2}(Q_k).
\end{equation}

Tables \ref{conv} and \ref{pmcdw} show that the way to achieve the renormalization scale independence is quite different for conventional and PMC scale-setting approaches. The scale independence under conventional scale-setting is due to the cancellation of the scale dependence among different orders, and the net scale uncertainty could be negligibly small by including enough high-order terms. It however cannot get precise value for each order. On the other hand, the scale independence of the PMC prediction is natural, which determines the optimal scale for each order by recursively using of RGE, thus it shall generally get scale-independent decay width at each order.

One can define a $K$ factor, $K=\Gamma_{\rm Total}/\Gamma_{\rm LO}$, which shows the relative importance of the high-order terms to the leading-order terms. Up to $\rm{N^4LO}$ level, the $K$ factor for $\mu_r = M_H$ under conventional and PMC scale-setting approaches have the following trends:
\begin{eqnarray}
K|_{\rm{Conv.}}\simeq1+0.53+0.07-0.04-0.02+\mathcal{O}(\alpha_s^5)
\end{eqnarray}
and
\begin{eqnarray}
K|_{\rm{PMC}}\simeq1+0.32-0.11-0.05+0.01+\mathcal{O}(\alpha_s^5).
\end{eqnarray}
Table  \ref{conv} shows that the perturbative nature of $K|_{\rm{Conv.}}$ changes greatly for different scales. For example, the $\rm{NLO}$ part of the $K$ factor changes by $\sim[-30\%,50\%]$ for $\mu_r\in[M_H/2,4M_H]$. On the other hand, Table \ref{pmcdw} indicates that if taking $\mu_r \neq M_H$, the perturbative nature of $K|_{\rm{PMC}}$ is almost unchanged.

\begin{figure}[htb]
\centering
\includegraphics[width=0.45\textwidth]{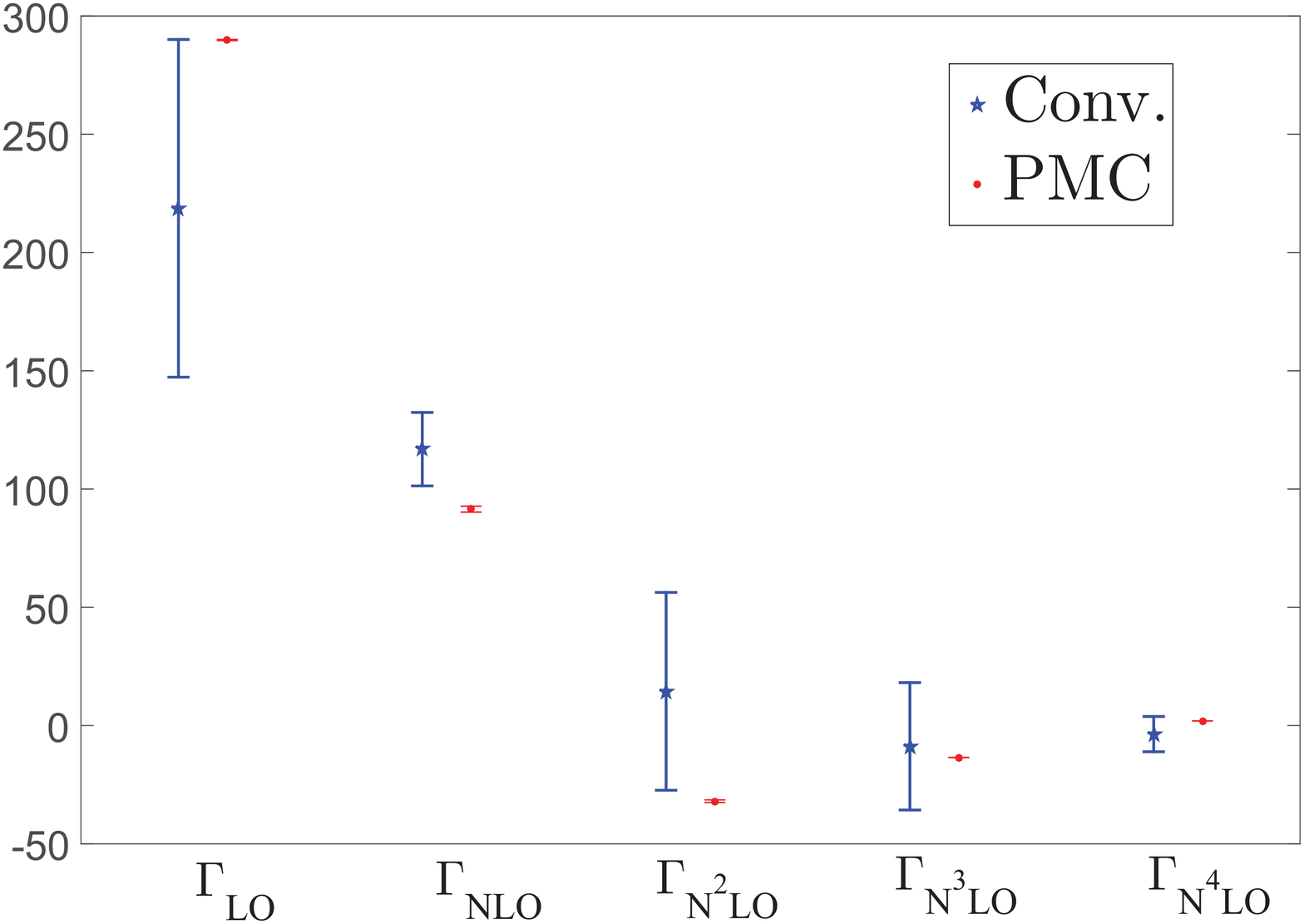}
\caption{Scale uncertainties of the individual decay width $\Gamma_i$ (in unit: $\rm{KeV}$) under conventional and PMC scale-setting approaches. $i =\rm{LO}$, $\rm{NLO}$, $\rm{N^2LO}$, $\rm{N^3LO}$, and $\rm{N^4LO}$, respectively. The central values are for $ \mu_ r = M_H $, and the errors are for $\mu _{r} \in \left[{M_H}/{2},4 M_H\right]$.}
\label{individual}
\end{figure}

More explicitly, we present a comparison of the scale uncertainties of the individual decay width $\Gamma_i$ under conventional and PMC scale-setting approaches in Fig.(\ref{individual}), in which the error bars are determined by
\begin{equation}
\Delta=\pm| \Gamma_{\rm{N^kLO}}(\mu_r)-\Gamma_{\rm{N^kLO}}(M_H)|_{\rm MAX}.
\end{equation}
Here the symbol `MAX' stands for the maximum value for $\mu _{r} \in \left[{M_H}/{2},4 M_H\right]$. Fig.(\ref{individual}) shows that the separate scale errors for each order are indeed quite large under conventional scale-setting, which are however negligible for all separate orders under PMC scale-setting.

\subsection{Residual scale dependence and an estimation of unknown high-order contributions after applying the PMC scale-setting}

\begin{figure}[htb]
\centering
\includegraphics[width=0.45\textwidth]{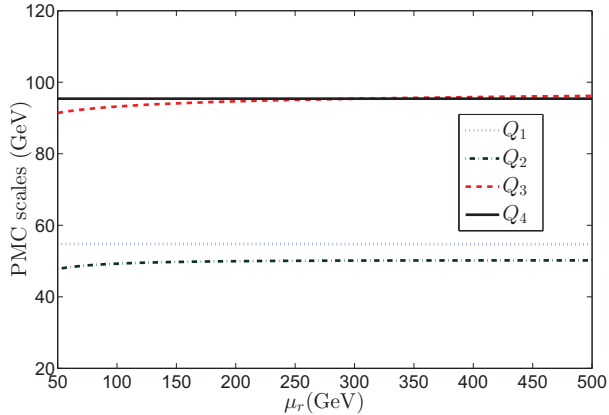}
\caption{The $\rm{LO}$, $\rm{NLO}$, $\rm{N^2LO}$ and  $\rm{N^3LO}$ PMC scales $Q_1$, $Q_2$ , $Q_3$ and $Q_4$ versus the initial choice of scale $\mu_r$ for $\Gamma (H \to gg)$ up to $\alpha_s^6$-order level, which are shown by the dotted line, the dash-dot line, the dashed line, and the solid line, respectively.}
\label{PMCmMOMQ}
\end{figure}

Eqs.(\ref{PMCQ11},\ref{PMCQ12},\ref{PMCQ13},\ref{PMCQ14}) show that the PMC scales $Q_i$ are of perturbative nature which absorb all the non-conformal $\{\beta_i\}$-terms of the pQCD series of $\Gamma (H \to gg)$ into themselves. If setting $\mu_r=M_H$, the perturbative series of the PMC scales behave in the following way
\begin{eqnarray}
\ln\frac{Q_1^2}{M_H^2} &=& -1.833+0.161+0.020+0.002+\mathcal{O}(\alpha_s^4), \label{Q1} \\
\ln\frac{Q_2^2}{M_H^2} &=& -2.419+0.731-0.163+\mathcal{O}(\alpha_s^3), \label{Q2} \\
\ln\frac{Q_3^2}{M_H^2} &=& -0.287-0.291+\mathcal{O}(\alpha_s^2), \label{Q3} \\
\ln\frac{Q_4^2}{M_H^2} &=& -0.542+\mathcal{O}(\alpha_s). \label{Q4}
\end{eqnarray}
Fig.(\ref{PMCmMOMQ}) presents the PMC scales $Q_{1,\cdots,4}$ versus the arbitrary initial choice of renormalization scale $\mu_r$. Those PMC scales are optimal and determine the correct behavior of the strong coupling at each order, all of which are unchanged for large values of $\mu_r$, e.g. $\mu_r \gg M_H$.

There are residual scale dependence due to unknown high-order terms, which however suffer from both the $\alpha_s$-suppression and the exponential suppression. Thus those residual scale dependence are generally small. Fig.(\ref{PMCmMOMQ}) shows the PMC scales $Q_{1,\cdots,4}$ are highly independent of the choice of $\mu_r$. By varying $\mu _{r} \in \left[{M_H}/{2},4 M_H\right]$, the values of those four PMC scales are almost unchanged:
\begin{eqnarray}
Q_1\simeq 55~{\rm{GeV}}, ~ Q_2 \simeq 50~\rm{GeV}, \nonumber \\
Q_3\simeq 94~{\rm{GeV}}, ~ Q_4 \simeq 95~\rm{GeV}.
\end{eqnarray}

\begin{figure}[htb]
\centering
\includegraphics[width=0.45\textwidth]{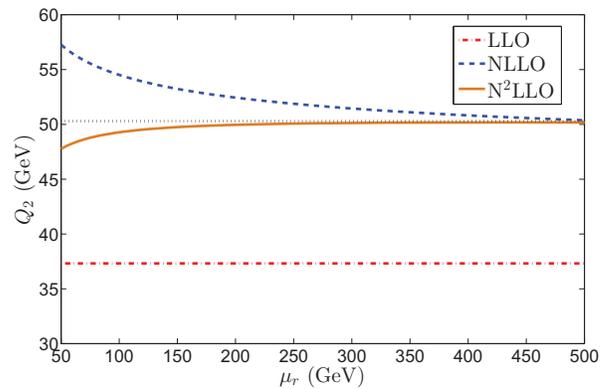}
\caption{A comparison of the PMC scale $Q_2$ from $\rm{LLO}$ accuracy up to $\rm{N^2LLO}$ accuracy. The dotted line is the approximate asymptotic limit for $Q_2$, which is for all orders. }
\label{PMCQ2}
\end{figure}

In our prediction on the decay width $\Gamma (H \to gg)$ at the $\alpha_s^5$-order level~\cite{Zeng:2015gha}, a somewhat larger residual scale dependence is existed for $Q_2$ at the $\rm{NLLO}$ accuracy. Fig.(\ref{PMCQ2}) shows how $Q_2$ changes when more loop terms have been taken into consideration. By using $Q_2$ up to $\rm{N^{2}LLO}$ accuracy, the residual scale dependence becomes much smaller. By varying $\mu_r\in[M_H/2,4M_H]$, the variation of $Q_2$ will be changed from $\Delta Q_2 \sim 7~\rm{GeV}$ for $\rm{NLLO}$ accuracy down to $\Delta Q_2 \sim 2~\rm{GeV}$ for $\rm{N^2LLO}$ accuracy. As a special case, it is found that the PMC scale such as $Q_2$ at the ${\rm LLO}$ accuracy shall be exactly free of $\mu_r$, a strict demonstration has been given in Ref.\cite{Brodsky:2013vpa}, which explains why the prototype of PMC, i.e. the well-known Brodsky-Lepage-Mackenzie (BLM) scale-setting approach~\cite{Brodsky:1982gc}, works so successfully in many of its one-loop applications. The PMC provides a underlying background for BLM and extend BLM up to all-orders by systematically identifying all of the $\{\beta_i\}$-terms at each perturbative order using a general ``degeneracy" pattern of the non-Abelian gauge theory~\cite{Bi:2015wea}.

\begin{figure}[htb]
\centering
\includegraphics[width=0.45\textwidth]{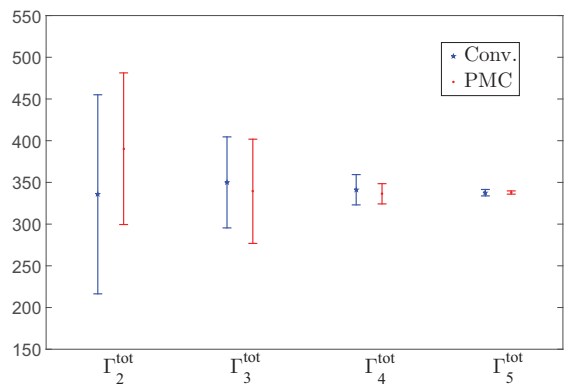}
\caption{Total decay width up to $k_{\rm th}$-loop level, $\rm{\Gamma_k^{tot}}$ (in unit: $\rm{KeV}$), for the decay width $\Gamma(H \to gg)$ together with the predicted unknown high-order contributions $\Delta \Gamma_k^{\rm tot}$ ($k = 2,\cdots,5$). The central values are for $\mu_ r = M_H $. }
\label{Gamma}
\end{figure}

As a final remark, it is helpful to be able to estimate the ``unknown" high-order pQCD corrections. The conventional error prediction obtained by simply varying the scale over a certain range is unreliable, since it only partly predicts the non-conformal contribution but not the conformal one. In contrast, after applying the PMC, the scales are optimized and cannot be varied; otherwise, one will explicitly break the renormalization group invariance, which leads to an unreliable prediction. We adopt the suggestion raised up by Ref.\cite{Wu:2014iba} for an estimation, i.e. for a $k_{\rm th}$-loop pQCD prediction for the $H\to gg$ decay,
\begin{eqnarray}
\Gamma_{k}^{\rm{tot}}=\frac{M_H^3 G_F}{36 \sqrt{2} \pi } \sum _{i=1}^{k}  C_i(\mu_r) a_{s}^{i+1}(Q_i[\mu_r]),
\end{eqnarray}
and the unknown high-order contribution is predicted by the following way
\begin{eqnarray}
\Delta \Gamma_{k}^{\rm{tot}}=\pm\frac{M_H^3 G_F}{36 \sqrt{2} \pi } \left|C_{k}(\mu_r) a_{s}^{k+1}(Q_k[\mu_r])\right|_{\rm{MAX}},
\end{eqnarray}
where $\mu_r\in[M_H/2,4M_H]$, and the symbol `MAX' indicates the maximum value. Under conventional scale-setting, $Q_i[\mu_r]\equiv\mu_r$; Under PMC scale-setting, $Q_i[\mu_r]$ are PMC scales and $C_{i}(\mu_r)\equiv r_{i,0}(\mu_r)$ are conformal coefficients. This way of estimating the unknown high-order pQCD prediction is natural for PMC, since after the PMC scale is set, the pQCD convergence is ensured and the dominant uncertainty is from the last term due to the unfixed PMC scale at this order.

Numerically, we obtain
\begin{eqnarray}
\Delta \Gamma_{2}^{\rm{tot}}|_{\rm Conv.} &=& \pm 119 {\rm KeV}, \;\;  \Delta \Gamma_{2}^{\rm{tot}}|_{\rm PMC}=\pm 91 {\rm KeV}, \\
\Delta \Gamma_{3}^{\rm{tot}}|_{\rm Conv.} &=& \pm 55 {\rm KeV}, \;\; \Delta \Gamma_{3}^{\rm{tot}}|_{\rm PMC}=\pm 62 {\rm KeV}, \\
\Delta \Gamma_{4}^{\rm{tot}}|_{\rm Conv.} &=& \pm 18 {\rm KeV}, \;\; \Delta \Gamma_{4}^{\rm{tot}}|_{\rm PMC}=\pm 12 {\rm KeV}, \\
\Delta \Gamma_{5}^{\rm{tot}}|_{\rm Conv.} &=& \pm 3.9 {\rm KeV}, \;\; \Delta \Gamma_{5}^{\rm{tot}}|_{\rm PMC}=\pm 1.9 {\rm KeV}.
\end{eqnarray}
A comparison of those results are presented in Fig.(\ref{Gamma}). The predicted error bars become smaller when more loop terms are included for both conventional and PMC scale-setting approaches, and the contributions from the unknown high-order terms become reasonably small.

\section{Summary}

In the paper, we have made a detailed analysis of the Higgs-boson decay $H \to gg$ up to $\alpha_s^6$-order. After applying the PMC, we obtain
\begin{equation}
\Gamma (H\to gg)\rm{|_{ PMC }} =337.9 \pm1.7_{-0.1}^{+0.9} \pm1.9 \;{\rm KeV},
\end{equation}
where the first error is caused by the Higgs mass uncertainty $\Delta M_{H}=0.24$ GeV, the second one is residual renormalization scale dependence for $\mu _r \in \left[{M_H}/{2},4 M_H\right]$, and the third one is the predicted unknown high-order contributions. To compare with the residual scale error $\left({}_{-2.4}^{+3.7}\right)$ KeV obtained by a $\alpha_s^5$-order prediction~\cite{Zeng:2015gha}, it is found that the residual scale dependence is greatly suppressed by including the newly calculated $\alpha_s^6$-order terms. Similarly, the scale uncertainty under conventional scale-setting shall also be suppressed by including the $\alpha_s^6$-order terms, which changes from the $\alpha_s^5$-order error $\left({}_{-8.1}^{+7.3}\right)$ KeV to a smaller error $\left({}_{-1.3}^{+2.1}\right)$ KeV.

Tables \ref{conv} and \ref{pmcdw} show that the scale dependence for the conventional and PMC scale-settings behave quite differently. The relatively smaller net total scale uncertainty for conventional scale-setting is achieved due to the cancellation of the scale dependence among different orders; thus even though the net scale uncertainty could be small by including higher-order terms, one cannot get precise values for each order by using the guessed scale. On the other hand, the net scale independence of the PMC prediction is rightly due to the scale-independence of all separate orders, since the PMC scale for each order is optimal and determined by properly using of RGE.

Our prediction on the Higgs-boson decay shows the conventional scheme-and-scale error could be avoided, which also emphasizes the importance of the renormalization scale-setting procedure after finishing the standard regularization and renormalization procedures. \\

\noindent{\bf Acknowledgement}: We thank the anonymous referees for helpful suggestions. This work was supported in part by the National Natural Science Foundation of China under Grant No.11625520, No.11547010 and No.11705033; by the Project of Guizhou Provincial Department of Science and Technology under Grant No.2016GZ42963 and the Key Project for Innovation Research Groups of Guizhou Provincial Department of Education under Grant No.KY[2016]028 and No.KY[2017]067.

\appendix

\section*{Appendix: New coefficients needed for a $\alpha_s^6$-order analysis}

The coefficients $c_{5,j}$ with $j=(0,1,\cdots,4)$ under the Landau gauge of the mMOM scheme are
\begin{eqnarray}
c_{5,0}&=& 1.17128\times 10^6 \ln ^4 {\frac{\mu_r^2}{M_H^2}}+1.73266\times 10^7 \ln ^3 {\frac{\mu_r^2}{M_H^2}} \nonumber\\
&&+5.88986\times 10^7 \ln ^2 {\frac{\mu_r^2}{M_H^2}}+2.08498\times 10^7 \ln {\frac{\mu_r^2}{M_H^2}} \nonumber \\
&&+\left(367840\ln {\frac{\mu_r^2}{M_H^2}}+411224\right)\ln ^2 {\frac{M_H^2}{m_t^2}}\nonumber \\
&&+ \Bigg (735680\ln ^2 {\frac{\mu_r^2}{M_H^2}}+2.11855\times 10^6 \ln  {\frac{\mu_r^2}{M_H^2}}\nonumber \\
&&+723634 \Bigg )\ln {\frac{M_H^2}{m_t^2}}+73568\ln ^3 {\frac{M_H^2}{m_t^2}}\nonumber \\
&&-4.8907\times10^7, \\
c_{5,1}&=& -283947\ln ^4 {\frac{\mu_r^2}{M_H^2}}-3.86754\times 10^6 \ln ^3 {\frac{\mu_r^2}{M_H^2}}\nonumber \\
&&-1.43599\times 10^7 \ln ^2 {\frac{\mu_r^2}{M_H^2}}-1.31706\times 10^7 \ln {\frac{\mu_r^2}{M_H^2}}\nonumber \\
&&+
\left(58666.7\ln {\frac{\mu_r^2}{M_H^2}}+73793.8 \right)\ln ^2 {\frac{M_H^2}{m_t^2}}\nonumber \\
&&+\Bigg(117333\ln ^2 {\frac{\mu_r^2}{M_H^2}}+367550\ln {\frac{\mu_r^2}{M_H^2}}\nonumber \\
&&+805.344\Bigg)\ln {\frac{M_H^2}{m_t^2}}+11733.3 \ln ^3 {\frac{M_H^2}{m_t^2}}\nonumber \\
&&+4.56382\times 10^6,  \\
c_{5,2}&=& 25813.3\ln ^4 {\frac{\mu_r^2}{M_H^2}}+309416\ln ^3 {\frac{\mu_r^2}{M_H^2}}\nonumber \\
&&+1.10391\times 10^6 \ln ^2 {\frac{\mu_r^2}{M_H^2}}+1.04041\times 10^6 \ln {\frac{\mu_r^2}{M_H^2}}\nonumber \\
&&+\left(-11164.4\ln {\frac{\mu_r^2}{M_H^2}}-2808.3\right)\ln ^2 {\frac{M_H^2}{m_t^2}}\nonumber \\
&&+\Bigg (-22328.9\ln ^2 {\frac{\mu_r^2}{M_H^2}}-32506.9\ln {\frac{\mu_r^2}{M_H^2}}\nonumber \\
&&-26489.6  \Bigg )\ln {\frac{M_H^2}{m_t^2}}-2232.89\ln ^3 {\frac{M_H^2}{m_t^2}}\nonumber \\
&& -137447,  \\
c_{5,3} &=& -1042.96\ln ^4 {\frac{\mu_r^2}{M_H^2}}-10315.9 \ln ^3 {\frac{\mu_r^2}{M_H^2}}\nonumber \\
&&-30130.1 \ln ^2 {\frac{\mu_r^2}{M_H^2}}-16904.8 \ln {\frac{\mu_r^2}{M_H^2}}\nonumber \\
&&+\left(379.259\ln {\frac{\mu_r^2}{M_H^2}}-192.395\right)\ln ^2 {\frac{M_H^2}{m_t^2}}\nonumber \\
&&+\Bigg(758.519\ln ^2 {\frac{\mu_r^2}{M_H^2}}-177.778\ln {\frac{\mu_r^2}{M_H^2}}\nonumber \\
&&+985.957\Bigg)\ln {\frac{M_H^2}{m_t^2}}+75.8519\ln ^3 {\frac{M_H^2}{m_t^2}}+5883.28, \nonumber\\
\\
c_{5,4}&=& 15.8025\ln ^4 {\frac{\mu_r^2}{M_H^2}}+115.885\ln ^3 {\frac{\mu_r^2}{M_H^2}}\nonumber \\
&&+183.216 \ln ^2 {\frac{\mu_r^2}{M_H^2}}-33.2589 \ln \frac{\mu_r^2}{M_H^2}-82.1927,\nonumber \\
\end{eqnarray}
where $m_t$ is top-quark pole mass.

The coefficients $r_{5,j}$, $j=(0,1,\cdots,4)$, under the Landau gauge of the mMOM-scheme are
\begin{eqnarray}
r_{5,0}&=& \frac{1}{48}\Bigg\{\left(1028979 \zeta_{3}-251939\right) c_{2,1}\nonumber \\
&&+3 \Bigg[761 c_{3,1}+437277 c_{3,2}-2568 c_{4,1}-84744 c_{4,2}\nonumber \\
&&-2097414 c_{4,3}+16 c_{5,0}+33 \Bigg(8 c_{5,1}\nonumber \\
&&+33 \left(4 c_{5,2}+66 c_{5,3}+1089 c_{5,4}\right)\Bigg)\Bigg]\Bigg\}, \\
r_{5,1}&=& \frac{1}{960}\Bigg\{\left(587874 \zeta_3+1817715 \zeta_5-278521\right) c_{2,1}\nonumber \\
&&-6 \Bigg[2 \left(1581 \zeta_3+7277\right) c_{3,1}+22 \left(4743 \zeta_3\right.\nonumber \\
&&\left.+37078\right) c_{3,2}-912 c_{4,1}-47430 c_{4,2}-1602909 c_{4,3}\nonumber \\
&&+48 c_{5,1}+1584 c_{5,2}+39204 c_{5,3}+862488 c_{5,4}\Bigg]\Bigg\}, \\
r_{5,2}&=& \frac{1}{1920}\Bigg\{12\Bigg[\left(48 \zeta_3+989\right) c_{3,1}+\left(7908 \zeta_3+77990\right) c_{3,2}\nonumber \\
&&+9 \left(-171 c_{4,2}-10979 c_{4,3}+4 c_{5,2}+198 c_{5,3}\right.\nonumber \\
&&\left.+6534 c_{5,4}\right)\Bigg]-\left(135672 \zeta_3+222720 \zeta_5\right.\nonumber \\
&&\left.+210167\right) c_{2,1}\Bigg\}, \\
r_{5,3} &=& \frac{1}{320} \Bigg\{-96 \left(3 \zeta_3-50\right) c_{2,1}-8 \left(48 \zeta_3+989\right) c_{3,2}\nonumber \\
&&+9 \Bigg[893 c_{4,3}-12 \left(c_{5,3}+66 c_{5,4}\right)\Bigg]\Bigg\}, \\
r_{5,4}&=&\frac{81 c_{5,4}}{80}.
\end{eqnarray}

\end{document}